%% file: main.tex
 \documentclass[manuscript]{acmart} 

\usepackage{array}
\usepackage{float}
\usepackage{booktabs}
\usepackage{tabularx}

\AtBeginDocument{%
  \providecommand\BibTeX{{%
    \normalfont B\kern-0.5em{\scshape i\kern-0.25em b}\kern-0.8em\TeX}}}

\setcopyright{acmlicensed}
\copyrightyear{2024}
\acmYear{2024}
\acmDOI{XXXXXXX.XXXXXXX}

\acmConference[Conference acronym 'XX]{Make sure to enter the correct
  conference title from your rights confirmation email}{June 03--05,
  2018}{Woodstock, NY}
\acmISBN{978-1-4503-XXXX-X/2018/06}

\setcopyright{none}
\renewcommand\footnotetextcopyrightpermission[1]{}
\makeatletter
\gdef\acmConference@shortname{}
\gdef\acmConference@name{}
\gdef\acmConference@date{}
\gdef\acmConference@venue{}
\makeatother
\AtBeginDocument{%
  \fancypagestyle{plain}{%
    \fancyhf{}%
  }%
  \fancypagestyle{standardpagestyle}{%
    \fancyhf{}%
  }%
  \fancypagestyle{firstpagestyle}{%
    \fancyhf{}%
  }%
  \pagestyle{empty}%
  \thispagestyle{empty}%
}

\title[Affordances of Gen Image AI for Design Communication]{Exploring the Affordances of Generative Image AI for Supporting Early--stage Architect--client Communication}

\author{Chengzhi Zhang}
\email{czhang694@gatech.edu}
\orcid{0000-0002-6868-2285}
\affiliation{%
  \institution{Georgia Institute of Technology}
  \streetaddress{}
  \city{Pittsburgh}
  \state{Pennsylvania}
  \country{USA}
  \postcode{15213}
}

\author{Weijie Wang}
\email{weijiew@alumni.cmu.edu}
\orcid{0009-0006-8467-2577}
\affiliation{%
 \institution{Independent Researcher}
 \country{USA}
}

\begin{abstract}

Text-to-image generative AI can produce renderings from natural-language prompts in near real time, making it increasingly popular for rapidly visualizing concepts in early-stage architectural design. Meanwhile, exchanging ideas efficiently and building shared understanding have long been central challenges in architect-client communication. How might the speed of generative image AI change this communication? To explore this question, we conducted a study with 11 architect-client pairs, in which each pair used generative image AI over video conference to collaboratively produce early-stage renderings of the client's "dream house." Our findings suggest that generative image AI helped pairs develop a solid shared understanding by providing concrete visual materials and supporting the exchange of ideas. It also shifted conversation dynamics, enabling clients to participate more actively in shaping design direction. However, challenges emerged, including a stylistic bias toward particular types of images and unpredictable shifts in design direction caused by variation across generations. We conclude with implications for the design of future generative image AI-based systems that support architect-client communication.
\end{abstract}



\begin{document}
\maketitle

\input{sections/1-introduction}
\input{sections/2-related_work}
\input{sections/3-recruitment}
\input{sections/4-activities}
\input{sections/5-findings}
\input{sections/6-discussion}
\input{sections/7-conclusion}

\begin{acks}
We thank Daragh Byrne, Nikolas Martelaro, and Paul Pangaro for their valuable input on this project, and the members of CoDeLab for their feedback throughout its development. This research was supported by the School of Architecture's Computational Design Research Micro-Grant. Finally, we are grateful to all study participants for their time and insights.
\end{acks}

\bibliographystyle{ACM-Reference-Format}
\bibliography{AIcomm}

\appendix

\input{sections/z-appendix}

\end{document}

%% file: sections/1-introduction.tex
\section{Introduction}

Effective communication is essential for successful design outcomes, while a lack of it can result in conflicting or misaligned goals~\cite{van2017good} and undermine design projects' success. Unlike the fine arts, where ``creative is pre-eminent, and clients, customers, users, and audience are not a consideration''~\cite{mcdonnell2011impositions}, architectural design centers on clients' needs, making effective communication crucial. However, subjectivity and expertise gaps can result in differences in stakeholders' understanding of the same linguistic symbol~\cite{chung2023artinter}, which challenges communication accuracy.


Researchers invented various communication tools to facilitate more effective communication~\cite{chiu2002organizational}. An architectural Information and Communication Technology (ICT) literature review~\cite{lu2015information} revealed the increase of ICTs during the 2010s. The boom brought tools including VR/AR systems~\cite{shiratuddin2011utilizing}, web-based systems~\cite{sun2010user}, Building Information Modeling (BIM) systems~\cite{gu2010understanding}, etc. These ICT tools have improved communication efficiency, but often created professional barriers for non-experts. After the 2010s, the ICT field experienced stagnation, revealing the need for new technologies to support communication. 

The rise of generative image AI offers a promising solution due to its ability to efficiently generate visual content, as evidenced by a recent review article~\cite{zhang2024research}. Compared with other design sub-domains, architecture is a field of heated research both in generative image AI~\cite{brisco2023exploring, koehler2023more, zhang_generative_2023, erouglu2022architectural, paananen2024using}, and ICT~\cite{lu2015information, shiratuddin2011utilizing, bates2012photorealistic}. Generative image AI is also referred to as text-to-image models (TTI)~\cite{chang_prompt_2023}, text-to-image generator (T2I)~\cite{chung2023promptpaint}, including prevailing products like Dall·E~\footnote{Dall·E: \url{https://openai.com/dall-e-2}}, Midjourney ~\footnote{Midjourney: \url{https://www.midjourney.com}}, Stable Diffusion~\footnote{Stable Diffusion: \url{https://stablediffusionweb.com/}} that can output high-fidelity~\footnote{Comparing with low-fidelity prototypes like paper and pencil sketches that are generally easy to create, high-fidelity prototypes like 3D models and renderings can represent designer's ideas accurately, but typically take more time to create~\cite{fay1990use}.} visuals in seconds with text input. Visual artists commonly believe that generative image AI can enable fast real-time communication~\cite{ko2023large}. A recent study in product design~\cite{he2024revealing} and a literature review in visual communication design~\cite{zhang2024research} further acknowledge the value of image AI in bridging the communication gap. These findings suggest that generative image AI has the potential to bridge design communication. 

In this study, we investigated how fast, high-quality renderings created by generative AI during architect-client conversations can facilitate/undermine design ideation and communication processes. To this end, we recruited 11 architect participants and client participants in pairs (22 participants in total) and invited them to use generative image AI (specifically, Stable Diffusion XL Beta Model on DreamStudio~\footnote{DreamStudio: https://dreamstudio.ai/generate} website) to collaborate on a design task while communicating with each other via video conferencing. Each design session is followed by a semi-structured interview with both participants. Our research questions are:


\begin{itemize}

    \item RQ1: How will early-stage one-on-one communication between architects and clients transform when both stakeholders adopt generative image AI?
    \item RQ2: What are the benefits and challenges of utilizing generative image AI in supporting synchronous architect-client one-on-one communication?
    \item RQ3: What are the implications for future collaborative systems design that incorporates generative image AI? 
     
\end{itemize}

Our preliminary findings with early-career architectural students as architect participants suggest that generative image AI helps quickly construct a ``common ground'' for communication by presenting high-fidelity visual materials. Client participants suggested that it enabled them to influence the design direction as they can also generate the images representing \textbf{their} desires. Meanwhile, we discovered some challenges and suggested implications for future generative AI-based system designs.

%% file: sections/2-related_work.tex
\section{Related work}

\subsection{Challenge of Architectural Design Communication}
  \label{section2.1}

Mertens et al.'s study has revealed the interaction of architects and end-users: during interaction and communication, the shared mental model is essential to the project's success~\cite{mertens2023interactions}. Adversely, ``architects are known to rely on their own experience as a main reference whilst designing~\cite{cuff1992architecture, imrie2003architects, verhulst2016whom}'', which makes linguistic communication unreliable, as each word (e.g., ``Japanese style'') carries drastically different meanings among architects. Apart from the abstractness in linguistic communication, the client's ``lack of grip in technical dimensions or specific language'' increases the communication gap~\cite{mertens2023interactions}. Consequently, clients~\footnote{We used the words ``client'', ``end-user'', and ``user'' interchangeably due to the fact that in architecture commissioning, clients are usually the ``occupants'' and ``users'' of the house. } without those language grips are inferior in communication. A lack of understanding of a client's needs can make architects seem ``arrogant''~\cite{angral2019architect}, which further undermines mutual communication.

Beyond architectural design, the ``fuzzy front end (FFE)'' stage is well recognized as an important stage across various design domains~\cite{varsaluoma2015fuzzy, opiyo2016approach, zhang2001fuzzy}. FFE is a stage involving a lot of idea generation, selection, and communication~\cite{koen2001providing}, with time creating low-fidelity prototypes and exchanging design ideas undermining communication efficiency. With this hardship, researchers propose user-involved approaches to enhance communication and requirement gathering~\cite{coughlan_effective_2002}, emphasizing the need for suitable methods and tools to support architects and clients as effective contributors~\cite{tzortzopoulos2006clients}.

\subsection{Practice for Assisting Architecture Design Communication}

Practitioners adopt various methods to overcome communication challenges. The most common analog practices are architectural design brief~\cite{bogers2008architects}, a common document to specify design requirements; design sketches~\cite{chandrasegaran2018sketching, self2019communication}, a commonly-used method for design ideation~\cite{prats2009transforming}. Compared with higher-fidelity 2D artifacts like schematics, the ambiguity in sketches promotes innovation~\cite{tseng2018can} while simultaneously introducing communication uncertainty~\cite{ekwaro2016uncertainty}. Thus, a medium combining both form and style is favored for communicating effectively.

Beyond those analog practices, various digital approaches have emerged with the rise of the architectural digital ecosystem~\cite{eckert1997intelligent}. Building Information Modeling (BIM) systems have been widely used to coordinate communication~\cite{oh2015integrated} by facilitating architect-client communication in the Schematic Design phase~\cite{tessema2008bim}. But BIM poses professional barriers to clients. Virtual reality (VR) was used to embody design outcomes and enhance collaborative design~\cite{frost2000virtual}, enhance architect-client collaboration~\cite{koutsabasis2012value}, and learner-centered communication~\cite{sopher2022exploring, tost2009worth}. However, the effort spent in developing the concept and modeling for the virtual world is tremendous~\cite{stempfle2002thinking}. Those existing practices either pose professional barriers to clients or are demanding, pushing us to explore new design communication tools.

\subsection{Generative Image AI as Visual Material}

In contrast with the abstractness in verbal communication mentioned in Section~\ref{section2.1}, images are concrete materials for various design contexts, and visual thinking is important for designers~\cite{ware2010visual}. This suggests that generative image AI could be a good tool for communication. Although generative 3D modeling appears more straightforward, it is still in its nascent stage as the generated meshes lack enough details~\cite{gao2022get3d}. Consequently, systems like~\textit{3Dall-E} rely on 2D image generation to inform 3D design and modeling processes~\cite{liu20233dall}. 

Generative image AI can also mediate design conversations by introducing ambiguity, unpredictability, and space for varied interpretations~\cite{yurman2022drawing}. It can further help support design ideation~\cite{brisco2023exploring, dortheimer2023think}, and inform physical sculpture-crafting~\cite{schroeder2023trash}. Recent systems include \textit{Artinter}, an art commissioning tool for gathering requirements and building mutual understanding between stakeholders~\cite{chung2023artinter}; \textit{Sketchforme}, an application for creating expressive and realistic sketches using text input~\cite{huang2019sketchforme}; and \textit{Opal}, a platform for generating news illustrations with news article input~\cite{liu2022opal}. This research reveals the promise of generative image AI as a resource for facilitating design conversations, as it allows rapid preparation of high-fidelity visuals in a short time with no professional barrier. 

%% file: sections/3-recruitment.tex
\section{Recruitment and Participation}

\subsection{Recruitment Procedure}

We recruited architect participants and client participants by posting flyers on [ANONYMIZED] university campus, using email lists and word-of-mouth methods. In recruitment materials (see Appendix \ref{appendix:recruitmentquestionnaire}), we screened architect participants with criteria: 1. Be proficient in English; 2. Have more than three years of design education experience; 3. Work in architecture or a closely related domain (like Urban Design, Computational Design, etc.) We screened client-participants with criteria: 1. Do not work or study in an architecture-related domain 2. Be willing to participate in this study as a client for a design project. Geographically, architect-participants were primarily from [ANONYMIZED] city, while client-participants were from different cities in the United States. After screening, we recruited 11 architect and client participant pairs. 

\subsection{Participants Demographic Information}

 Table \ref{table:designerdemographic} and Table \ref{table:clientdemographic} provided demographic information of architect and client participants. Architect participants averaged 7.18 (SD = 1.83) years of design education. Although most architect participants reported themselves as master students (N = 8), they all had at least \textbf{1 year of work experience}, averaging 1.86 (SD = 0.84) years of work experience, comparable to early-career architects. We did not intentionally recruit a large portion of computational design (a sub-domain within architecture design) majors, while computational design majors are more familiar with computer-assisted design practice, which enables them to use the generative image AI tool. Client participants' ages averaged 26.27 years old (Min = 23, Max = 30, SD = 2.65). Architect (A) and client (C) participants are matched by their participant number (i.e., A1-C1, A2-C2,..., A11-C11). In Table \ref{table:designerdemographic}, the ``Hours spent on generative image AI'' implies the self-reported total hours spent on any generative image AI software before the study.

\begin{table}
\caption{Demographic information for architect participants} 
\label{table:designerdemographic}
\resizebox{\textwidth}{!}{%
\begin{tabular}{llllllll}
\hline
\textbf{ID}  & \textbf{Age} & \textbf{Gender}& \textbf{Major/ Job Title}   & \begin{tabular}[c]{@{}l@{}}\textbf{Design Student/}\\ \textbf{Professional}\end{tabular} &\begin{tabular}[c]{@{}l@{}}\textbf{Hours Spent} \\ \textbf{on Generative} \\ \textbf{Image AI}\end{tabular} & \begin{tabular}[c]{@{}l@{}} \textbf{Work}\\ \textbf{Exp.}\\ \textbf{Years}\end{tabular} & \begin{tabular}[c]{@{}l@{}} \textbf{Study} \\ \textbf{Exp. }\\ \textbf{Years}\end{tabular} \\ \hline
A1  & 24  & F      & Computational Design    & Design Student/Masters  & 1-5  & 1 & 6   \\  
A2  & 28  & M      & Computational Design    & Design Student/Ph.D.  & 5-10 & 1-2 & 8-10 \\  
A3  & 27  & M      & Computational Design    & Design Student/Masters    & 0-1 & 3   & 6  \\  
A4  & 33  & M      & Computational Design    & Design Student/Masters  & 5-10 & 3   & 10  \\  
A5  & 25  & M      & Urban Design   & Design Student/Masters  & 0-1 & 1  & 7 \\  
A6  & 25  & M      & Computational Design    & Design Student/Masters  & 1-5  & 1   & 7 \\  
A7  & 25  & F      & Computational Design    & Design Student/Masters   & 1-5 & 2  & 7  \\ 
A8  & 25  & F      & Building Performance & Design Student/Masters    & 0-1 & 1 & 7  \\  
A9  & 28  & F      & Computational Design    & Design Student/Ph.D.   & 1-5 & 2  & 10   \\  
A10 & 25  & M      & Computational Design    & Design Student/Masters  & 5-10 & 3  & 6  \\  
A11 & 24  & F      & Architectural Assistant & Professional Designer   & 0-1 & 2  & 4  \\ \hline
\end{tabular}%
    }
\end{table}

\begin{table}[h]

\caption{Demographic information for client-participants}
\label{table:clientdemographic}

\begin{tabular}{ccccc}
\hline
\textbf{ID}  & \textbf{Age} & \textbf{Gender} & \textbf{Student/Professional} & \textbf{Major/ Job Title}                \\ \hline
C1  & 23      &   F & Masters Student     & Human-Computer Interaction       \\ 
C2  & 25     &  M & Ph.D. Student    & Mechanical Engineering           \\ 
C3  & 26     &  M  & Ph.D. Student      & Human-Computer Interaction       \\ 
C4  & 25     &  F & Professional   & Design                           \\ 
C5  & 25     &  F & Professional     & Interaction Designer             \\ 
C6  & 24     & F & Professional    & Data Scientist                   \\ 
C7  & 30     &  F & Ph.D. Student   & Learning Sciences                \\ 
C8  & 24    & M   & Ph.D. Student  & Computer Science and Engineering \\ 
C9  & 24    &  M   & Ph.D. Student   & Computer Science                 \\ 
C10 & 26    & F  & Masters Student    & Industrial Design                \\ 
C11 & 24   &M   & Masters Student   & Digital Media                    \\ \hline
\end{tabular}

\end{table}

%% file: sections/4-activities.tex
\section{Study Organization and Activities}

\subsection{Study Activities}

Architect-client pairs were invited to participate in an 80-minute session over video conferencing. Each participant was compensated with a USD \$15 Amazon gift card for their time. We deliberately chose a \textbf{remote} study setting to ensure both participants can have equal access to image generation and image sharing, without having one party taking full control at a time.

The study consisted of a 10-minute introduction section, a 45-minute design section, and a 20-minute interview section. During each study session, architect and client participants used Zoom video conferencing~\footnote{Zoom: https://zoom.us/} and collaborated on a shared online whiteboard (Figma~\footnote{Figma: \url{https://www.figma.com/about/}}) using a template (see Fig.~\ref{fig:figmaboard}) we provided. On the template, there are squares (image boxes) for filling in generated images and a text box below each square for entering the corresponding prompt. Architect and client participants used DreamStudio (with the Stable Diffusion XL Beta API) \footnote{DreamStudio: \url{https://dreamstudio.ai/generate}} by logging into the research account we provided. Participants used the DreamStudio independently to create variations/rendering. 

\subsubsection{Introduction Section}

Before the study, all participants signed a consent form approved by our institution’s IRB. The 10-minute introduction section is designed to familiarize people with the image generation and collaboration workflow. We began with a review of the onboarding document (see Appendix \ref{appendix:onboardingdocument}), and then participants were given time to familiarize themselves with DreamStudio. They were asked to practice by generating images of a sofa and copying and pasting one generated image to the Figma board along with the prompt they had used (e.g., \texttt{A sofa with some classical ancient Chinese styles}).

\subsubsection{Design Section}

During the design section, participants worked collaboratively on the Figma board and communicated via Zoom conferencing. To start the design section, we assigned the design task as follows:

\begin{quote}
The client participant asks the designer participant to design a house by the sea for the client. Please use the design section to achieve this goal together.
\end{quote}

\begin{figure}[h]
  \centering
  \includegraphics[width=1\textwidth]{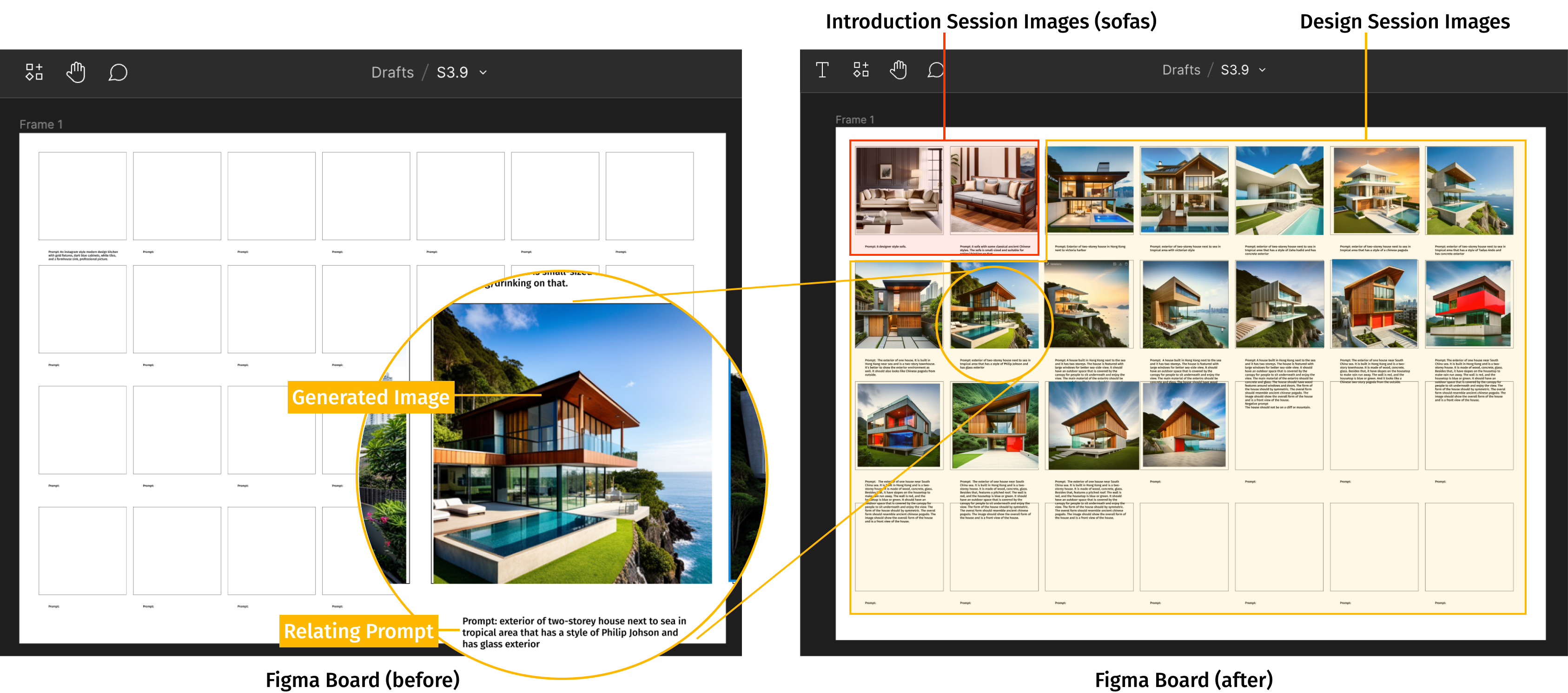}
  \Description[Two Figma boards before and after the study]{The left Figma board consists of black squares and text ``prompt'' beneath each black square. The right Figma board consists of many images filled inside the square, and beneath each image, there is a long text description.}
  \caption{Image of the Figma Board Before (left) and After (right) the Design Section}
  \label{fig:figmaboard}
\end{figure}

We chose the ``dream house by the sea'' task to set the client participant in the anticipated ``architect-client'' communicative scenario, as anyone can have their ``dream house'' and personal preferences, while the ``by the sea'' requirement introduced a constraint so we could compare against each session. Participants were also allowed to use any software/search engines needed. They were asked to work iteratively on the design concept until either they were satisfied with the design and wished to stop, or they reached the time limit of 45 minutes. A typical Figma board before and after the session is shown in Fig. \ref{fig:figmaboard}.

\subsubsection{Interview Section}

Immediately following the design task, we proceeded to a semi-structured interview. This was completed with the architect participant and the client participant individually. Each participant was interviewed for approximately 10 minutes.

\subsection{Qualitative Analysis}

We used Otter.ai \footnote{Otter.ai: \url{https://otter.ai/home}} to transcribe interview section videos into texts and then analyzed all interviewees' responses. Researchers also referred to the original audio recording when the transcript was inaccurate and refined the transcript accordingly. We adopted thematic analysis for developing themes~\cite{braun2006using}.

The analysis process consisted of two rounds--initial coding and refined coding. For the initial coding, the first author reviewed all of the transcripts and familiarized themselves with the transcripts, then developed initial codes that represent the patterns in the data. The second round of coding involved the first and second authors. Each author individually reviewed the transcripts and coded them using the initial codes, making adjustments as needed. Excerpts that have similar meanings have the same codes applied. The first and second authors used affinity diagramming for developing emerging themes while discussing with one another. 

%% file: sections/5-findings.tex
\section{Study Findings}

\subsection{RQ1: What Works Well and Benefits to Communication}

\subsubsection{Generative Image AI Provides Solid Visual Materials} 

While the terms they used varied and included ``reference images'' (A1) or ``get what you want in the form of a picture'' (C11), participants generally appreciated the generative image AI’s role in providing visual materials for assisting discussion (A1, A2, A3, A4, C4, C5, C7, C9, A10, C11). Those reference images provide a general sense of what looks good to participants, and are the visual materials they can ``directly point to'' (A3). Such aspects helped to reach a rough shared understanding of the design concept. C2 stated: \textit{``AI is being a huge part like ... \textbf{representing my ideas}, even though I'm giving him [the AI] so little.''} A9 stressed that visual assistance is important, and generated images help avoid misunderstanding and miscommunication. Just as C4 stated: \textit{``What works well is to \textbf{have something to talk about}, rather than trying to think off the top of your head.''} The generated images often did not provide the ``perfect'' result that would represent what participants ideally wanted (A2). Still, client participants favored generating their own images because the architect participants could consider their opinion by looking at what \textbf{they} generated (C3, C4, C5). C5 noted: \textit{``AI will visualize the part--maybe you like it--so you can tell the designer that this is exactly what I want.''} C4 added that the generated images could complement traditional design specification documents that contain quantitative data like budget, cost, and room numbers.

\subsubsection{Generative Image AI Facilitates More Efficient Communication}

Many participants noted that generative image AI promotes communication by making it more efficient (A1, C2, C6) and faster (A1, C1, C4, C6, A9). For instance, A1 stated, \textit{``AI is really good for fast results.''} A2 mentioned that image AI helps save a lot of communication effort towards mutual understanding: \textit{``this step actually \textbf{saves a lot of, a lot of time} like understanding each other.''} Participants were satisfied with both the process (C3) and results (A4, C10). A4 stated: \textit{``I feel it greatly increases the process of communication... they increase their \textbf{work efficiency and communication efficiency}.''} A10 also described the images as supporting faster decision-making, as \textit{``It's quick, very intuitive to see, okay, this is what I like, and what I don't like... it can make that decision process \textbf{much faster}.''}

\subsubsection{Participants Think AI Sparks Design Inspirations and Provides Unique Designs} 

Participants recognized AI's role in providing design inspiration that is helpful for early-stage design. Both architects and clients value the fresh perspectives and ideas generative image AI brings to the table (A4, A8, C8, A9, C11). C11 noted: \textit{``it can be used as the inspiration for the beginning point.''} A4 added: \textit{``Dall-E \footnote{ This refers to DreamStudio, not DALL·E. The confusion likely stems from DALL·E's widespread recognition as a leading AI image generator.} generates some unexpected 2d image ... that definitely \textit{gives some inspirations}.''} Comparing the generated images with those found on the internet, participants regarded generated images as providing more ``unique'' design solutions (C7). C11 preferred AI-generated images as they give ``directly what you want'': \textit{``you cannot search like this on Pinterest or something, it cannot give you the picture exactly aligned with what it described there.''}

\subsubsection{Image AI Refines Clients' Desires, Supports Decision-making, and Clarifies Preferences}

Architects suggested that the generative image AI transformed vague or undefined client desires into a concrete understanding of their preferences. C11 stated: \textit{``clients may not always know what they exactly want''}, and highlighted how AI helps find out clients' desires. C2 highlighted~\textit{``It helps me a great deal by \textbf{sorting out} that you probably want a great layout wall... and you probably have a huge window across. That's pretty good.''} A4 and A8 thought image AI helped clients figure out their true needs: \textit{``This helps the client to figure out what they really want''} (A4). The generated images have helped clients recognize, interpret, and uncover their true desires as they can see the various possibilities and know their own preferences more precisely.

\subsubsection{Prompt Sharing Processes Benefit Both Architects and Clients} 

In our study, the clients used the architect’s prompts more often than the other way around, as architects often use more specific design terms and language descriptions. Client participants appreciated prompt sharing processes (C3, C4, C7, C9, C11). Architect participants articulated some professional terms that client participants can use to adapt and refine their prompts (e.g., \textit{``modernism and post-modernism''} that C4 mentioned). C5 remarked: \textit{``I definitely copied some of my designer's prompts and tried to \textbf{change the keyword} to see whether he can achieve what I want.''} Architect participants also draw key information after communicating with the client participant to construct proper prompts as inputs (A8). Participants wished to have better collaborative prompt engineering tools (C8, C11), while A8 preferred having a list of prompts for assisting prompt engineering processes. 

\subsection{RQ2: What Did Not Work Well and Challenges to Communication}
\label{section:5.2}

\subsubsection{Participants Find It Hard to Generate Images They Want} 

Participants felt a lack of control over the generated images (C1, A1, C2, A3, A6, A7, C8, A8, C9, A11) and complained \textit{``having no control is the biggest issue with these AI''} (C1), \textit{``there's very little control we have over the results''} (C2), and \textit{``I don't feel I have much control over the image''} (A11). Image AI did not follow what they specified in the prompts (A6) and ignored some specified prompts (C3). A5 was confused whether it was the software or the architect who was wrong in this process. C8 mentioned: \textit{``the results sometimes \textbf{differ too much} from what I wanted. And there's no way to change it or \textbf{steer into the way I wanted}.''} We observed an inappropriate ``instructional'' prompts (prompting AI as if it is an agent, e.g.,\texttt{``Please put the desk to the center of the room''}) instead of the ``descriptional'' prompts (directly describing what should be on the image, e.g., \texttt{``Exterior of a two-story house in Hong Kong next to Victoria Harbor''}). 

\subsubsection{Participants Feel AI is Unable to Meet Specific Design Requirements} 

Participants find it hard for AI to generate design results that meet their specific rule-based design requirements (A2, C2, C5, A7). For instance, A2 and C2 specified they wanted a four-floor house, but failed as AI consistently generated houses with 2 or 2.5-ish floors. A8 found it impossible to generate the architecture with a set number of floors and a specific color of facade. When working on interior images of a home, C5 specified in the prompt to put a table in the corner of a room, but AI kept generating more ``common'' layouts having a coffee table at the front of the sofa. A7 shared that the designs generated did not correspond to a set budget, as those images were ``super fancy.'' Generally, our participants found it extremely hard to generate highly specific designs or achieve detailed design requirements. 

\subsubsection{Perceived Bias in Generated Images} 

\begin{figure}[h]
  \centering
  \includegraphics[width=0.8\textwidth]{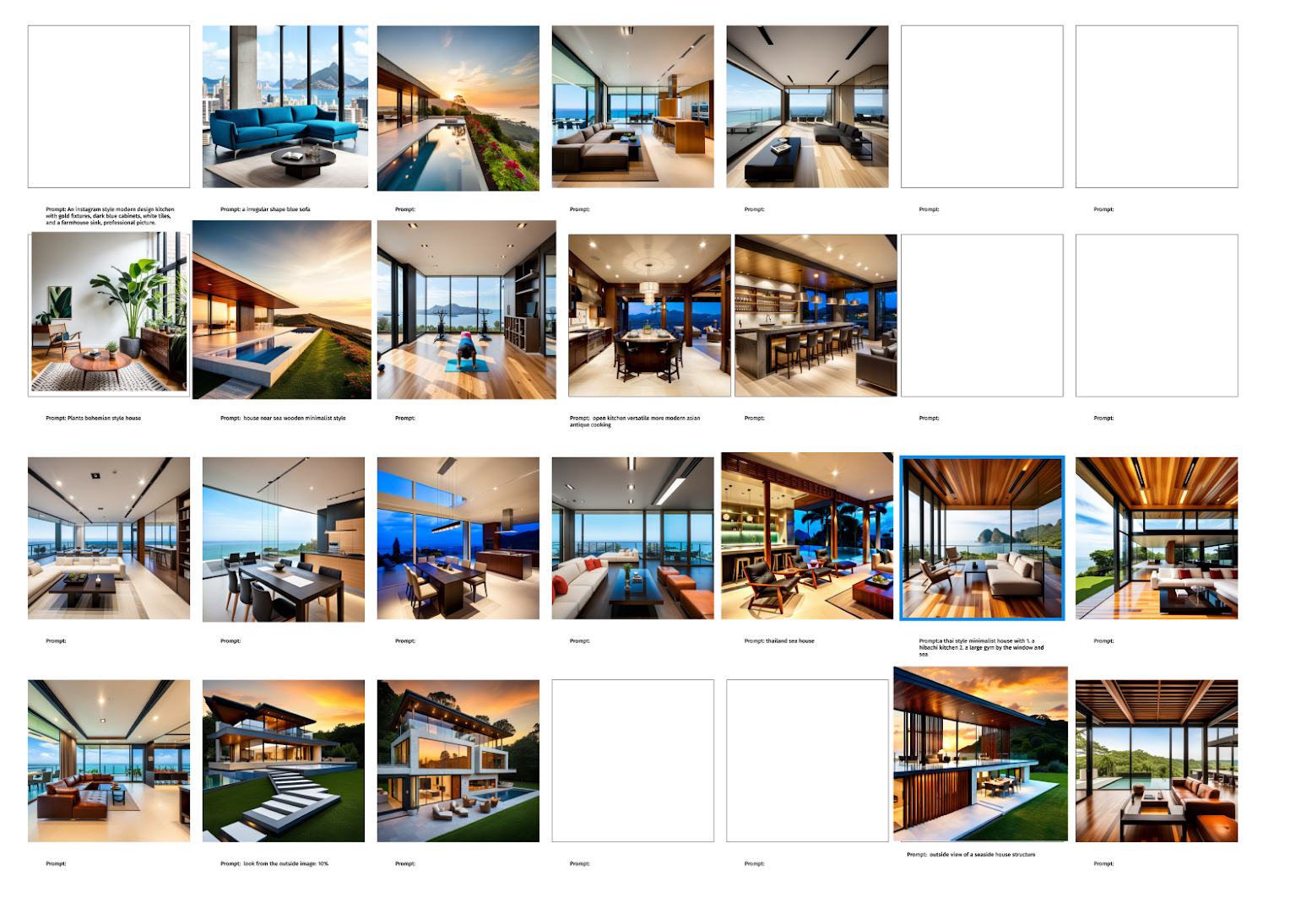}
  \Description[Many highly saturated fancy images]{Many images depicting the interior and exterior of different fancy houses. Beneath some images, there are some text descriptions}
  \caption{Screenshot of study 7 Figma board, where participants identified a tendency towards ``high-end'' style in generated images}
  \label{fig:bias_images}
\end{figure}

Many participants mentioned that DreamStudio tends to output images that look ``dreamy''(A4), ``fancy looking'' (A7), similar-looking (A2, A10), and with high saturation (A9), which deviates from their intent and preference (see Fig. \ref{fig:bias_images}). C5 witnessed the bias by stating \textit{``AI always generated some image with a \textbf{high probability}, but it's not the concept for myself.''} Likewise, C1 stated: \textit{``AI tool is very \textbf{limiting}, it tends to have these like very \textbf{popular images}.''} A7 thought AI-generated images seem like exhibition images rather than realistic designs, hindering communication accuracy. AI also failed to produce images in some specific design styles, as C9 wanted to generate a ``Chinese style'' exterior but did not get what they wanted, even if specified in the prompt--\textit{``It should also look like a Chinese pagoda from outside.''} AI's generating popular styles limited participants from exploring design possibilities and constrained their creativity. 

\subsubsection{\textit{``Architecture is Reduced to 2D Images''}} 

A clear limitation of 2D generative images is that sometimes people can not interpret a feasible architectural design scheme from the generated images (A8), as those images may not translate correctly to a feasible architectural layout design. As each generation creates completely new images, it is also impossible to generate other perspectives of a specific architectural design (A1, C1, C10). Some images did not make sense in reality at all (C3). A11 worried that the generated images reduced architecture to images, losing the materiality and context: \textit{``... the challenge facing us here is that \textbf{architecture is reduced to images}... And in reality, I think architecture is more about a presence in time and space. So it is a \textbf{three-dimensional experience}, which will carry information about materiality and structure and also perhaps the location.''} C11 was also concerned that it reduced architectural work to the processes of verbalization and generation. 

Alternatively, A4 expressed a preference for generative 3D due to the implicit limitations of designing architectural forms through 2D images. They also voiced their concern about a loss of agency, as the design process should be progressive, while AI just provided design decisions, lacking the potential for further development: \textit{``I lost all the sense of design. I don't know what to do next. I feel like the design was finished in a second, and the results are totally not as I expected because it's not my design.''} Architects thought it was not professional to incorporate AI in design processes (A8), as they could not anticipate the outcomes (A3), nor could they utilize professional thinking and skills (C11). A9 suggested that AI led to the loss of the architect's style, stating: \textit{``... within each of the hand drawings, or their idea generated by the person, it has the designer's personality in it. Right now, the tool does not have any personality.''}

\subsubsection{Design Directions Shift Too Rapidly Between Rounds of Generation} 

\begin{figure}[htbp]
  \centering
  \includegraphics[width=0.8\textwidth]{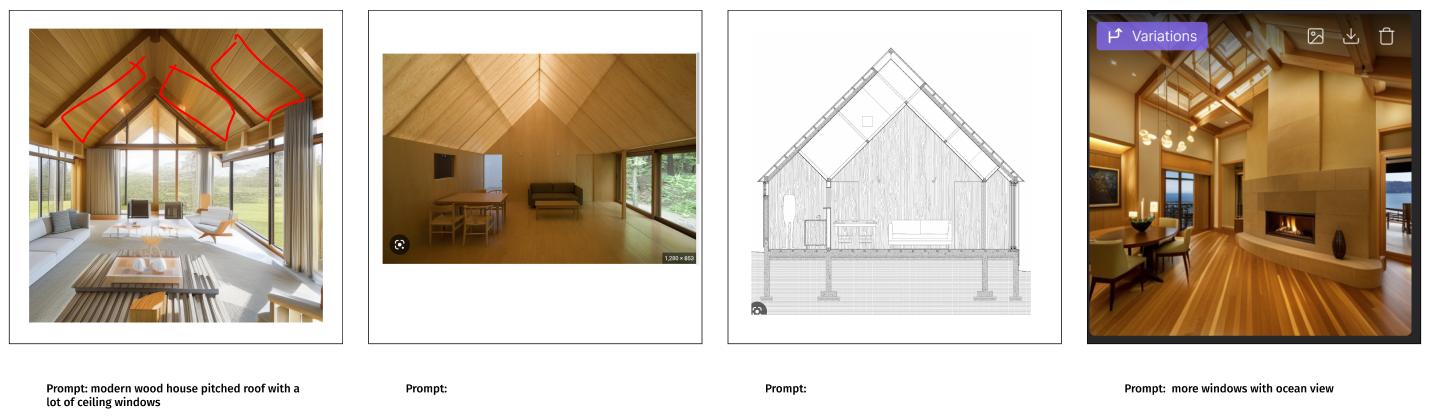}
    \Description[Four images of rooftop design]{Four images of rooftop designs in parallel with each other, with the first one, the second one, and the last one in rendering style. The third one is in sketch style. Within three rendering style images, the interiors and the rooftops look dramatically different.}
  \caption{Screenshot of study 7 Figma board, where participants identify abrupt shifts in design solutions (with the leftmost and rightmost images generated subsequently but with totally different looks)}
  \label{fig:directions_shift}
\end{figure}

Participants noted that with each round of generation, the design is totally different (A1), and the progressive and iterative nature of the design process is compromised (A5). In A5's words: \textit{``Like, I can give you \textbf{something new but not totally new} so that you can have a back and forth. But when we are using this software, there's no back and forth. It's totally direct, like, it's just it. \textbf{Boom, that's it.}''} Even with the same prompt, the outcome is different each time (A10, C10), lacking progressive improvement processes (A1). It may not be obvious to the client participant, but from the designer participant's view, each image represents a distinct design direction. For example, A3 regarded the left-most and right-most images as totally different designs as they are different structures, but C3 regarded them as similar designs as they both have pitched roofs (see Fig. \ref{fig:directions_shift}). Architect participants complained that client participants can easily be distracted by different design choices (A3): \textit{``...image generator can just change it every second, it changes so fast and architects \textbf{lose their track of thoughts}''}. 

\subsection{RQ3: Improvement Direction for Incorporating Generative Image AI in Design Processes}

\subsubsection{Participants Seek More Precise and Flexible Control over the Outcome} 

 Participants desired more control and input flexibility (A2, C2, C9, A11), as now image generation was like using a slot machine, in which participants ``hoped'' they could get good results, but many times in vain. Possible solutions could be setting some hard-coded rules for AI image generation (C2, C9), using sketches as input (C1, C3), and partial generation (A3). C2 mentioned, \textit{``in reinforcement learning, you have more \textbf{hard-coded} or like stronger restrictions''}. A3 mentioned that \textit{``If I want to remove the pool on the second floor, maybe I can circle that area and... type something like remove the pool or replace the pool with some other elements.''} Partial generation (which is known as in-painting) is a tool that we already see in tools like Dall-E 2.

\subsubsection{Participants Prefer Conversational Interaction} 

Compared with ``describing'' what should be on the image in the prompt, participants favor using conversational user interfaces interacting with generative image AI, like directly editing images using verbal input (D1, C4). A1 mentioned that \textit{``we uploaded the image and gave prompts to modify it.''} Another suggestion from C6 is that users could have a conversation with AI, like AI agent could ask users some questions to proceed further: \textit{``It would be better if it could ask me some questions like we can have a conversation and then AI generates the image ... if AI needs more clarification, it should ask me and give me some options.''} Rather than using generative image AI as a ``rendering'' tool, participants generally preferred conversational input for understanding their preferences in a more natural way.

\subsubsection{Facilitating User Personalization and Divergent Thinking}

Some participants suggested that the AI model could be customized and learn from their input. C9 stated \textit{``maybe DreamStudio itself can train based on our selection and perform better in the future.''} C11 also preferred that \textit{``it can learn from what you have previously fed into the machine''}, similar to having reinforcement learning embedded within the generative system. Participants discussed having more diversity in the AI-generated images. C3 mentioned that they wanted divergent thinking sometimes, but DreamStudio did not support that process. They stated: \textit{``Maybe it needs to have a function where I can just look for some more like \textbf{dramatically different styles} so they'll fit into some of the criteria I was interested in.''}

\subsubsection{Supporting Collaborative Prompt Engineering and Image Co-creation}

Participants favor an all-in-one platform in which they can streamline the generation-sharing processes (C11). The current copy-paste process to take images from the generative AI and place them into the shared Figma whiteboard requires extra work. Alternatively, participants suggested that a shared, generative workspace could be an improvement direction. C11 favored having an integrated platform to facilitate communication and improve each other's prompts by saying \textit{``maybe there should be a way for people to collaboratively use prompts or collaboratively share prompts and work on prompt modification together, something like that.''} 

%% file: sections/6-discussion.tex
\section{Discussion}

\subsection{Generative Image AI Altered Design Communication}

Both client participants and the architect participants from our study valued the high-resolution design renderings generated by generative image AI, which were based on and relevant to the prompt input. Those visual materials synthesize that which language can not adequately convey, mitigating the vagueness and ambiguity of linguistic communication~\cite{de2020multimodal, chung2022artist}. As a result, the stakeholders do not need to \textit{``think off the top of their heads''} (C4); instead, images could help them be~\textit{``on the same page''} (C2) in an effective way.

Beyond the effectiveness in bridging communication gaps, the approachability of generative image AI was also favored. Traditional mediums such as building information modeling (BIM) systems~\cite{tessema2008bim} and virtual reality (VR)~\cite{koutsabasis2012value} often require significant professional expertise and design effort. In contrast, generative image AI is more accessible and user-friendly, even for participants with no prior experience, including A3, A5, A11, C5, and C6. Output-wise, image AI advances over traditional communication practices like sketches or reference images (which can lack relevance either in form or style). Generated images are often photo-realistic, intuitive, and easy to interpret for both stakeholders.

Many participants valued generative image AI for bringing new ideas and perspectives (A4, A7, A9, C8, C11), likened to ``provocateur'' in the design processes~\cite{steinfeld2021significant}. This finding aligned with previous studies that see image AI as capable of supporting creativity, reducing cognitive load~\cite{chandrasekera2025can}, unleashing imaginations, and ``rectifying humans' biased creation''~\cite{ko2023large}. However, existing studies also revealed that the use of AI-generated images could lead to design fixation~\cite{wadinambiarachchi2024effects}. Although designing for a client participant's specific preferences and designing for ideation (which study witnessed the design fixation) are distinct tasks, future rigorous evaluation could help determine whether introducing AI truly inspires participants' imagination in a more refined design context, or if it merely creates a false perception of inspiration.

\subsection{Generative Image AI Changed Power Dynamics and Beyond}

Generative image AI changed the power dynamics between client participants and architect participants. Client participants noted that generative image AI empowered them to engage in more meaningful discussions with designers by providing visual materials that reflected \textbf{their} ideas and goals (C3, C4, C5). Previously, their lack of design expertise limited their ability to create visual materials---such as sketches, or other materials discussed in the related work section---that could effectively communicate their vision. Generative AI provided them with greater voice, agency, and efficiency, enabling them to participate more actively in the design process.

Clients often expressed a desire to be more involved in the design decision-making process~\cite{kilinc2015changing}, while their limited domain-specific terminology and the lack of professional skill set significantly hinder their full participation~\cite{mertens2023interactions}. Addressing this challenge, we observed that using generative image AI, prompting collaborative, and creating 2D ``design'' effectively reduced these barriers. During their communication and prompt-sharing processes, client participants actively picked up terms (like ``modernism'', ``post-modernism'') from architect participants, and subsequently used these terms to generate \textbf{their} own ideas. This altered the traditional relational structure, in which clients were often relegated to passive receivers of the architect's ideas. The whole process suggested a more reciprocal architect–client relationship~\cite{siva2011investigating}. 

Some architect participants were critical of this dynamic shift. They worried that it deviated from the traditional progressive architectural design practice that typically worked from ``simple volumetric definitions to gradually more detailed design''~\cite{joyce2021ai}. A3 regarded such results unpredictability as unprofessional, and A11 expressed that architecture is a presence in time and space, a lack of those considerations deviated from the essence of architecture that \textit{``Architecture is reduced to 2D images.''} A8 also expressed worries about AI replacing their domain expertise. In an era when AI is evolving the role of creative practitioners~\cite{palani2024evolving}, such critical socio-technical aspects should be considered for human-centered generative AI systems. For instance, some design processes, such as building empathy with clients, can not be simply automated with AI-generated results~\cite{lu2024ai}. The perceived efficiency gains in AI-mediated communication should be weighed against hidden risks to best preserve the essence of design in this AI era.

\subsection{Implications for Future Generative Image AI Systems Design}

In addition to the opportunities for improved design communication, the study findings suggest future directions for systems incorporating generative image AI.

\subsubsection{Improving User Control with Symbolic-connectionist System}

Even with well-designed prompts, generative AI may or may not adhere to all the texts specified in the prompt, causing the text-image alignment problem~\cite{saharia2022photorealistic}. Participants expressed the need for more flexible and precise control, as well as greater controllability over the image outcome~\cite{ko2023large}. Existing research uses multi-modal input and sketch input to define the shape and steer the results~\cite{chung2023promptpaint, zhang2023adding, lin2025inkspire}. Beyond the mentioned approaches, we identified the need for symbolic-connectionist AI systems, as participants mentioned they wanted to set some ``rules'' for AI to act on (C2, C9). This is particularly important because the architecture domain often demands high levels of specificity, such as the number of floors and rooms. Unlike symbolic AI, which functions by processing and reasoning over human-interpretable symbols, generative image AI often uses connectionist AI algorithms. Connectionist AI relies on distributed representations and learns patterns from data, making it difficult to follow specific text input rules~\cite{mollo2023vector}. The need also aligned with symbolic-connectionist systems proposed by researchers like S. Harnad \cite{harnad1990symbol}: connectionist approaches to AI (e.g., deep neural networks) would be better equipped to provide necessary iconic and categorical representations (i.e., the ``rules'' participants mentioned). Future generative image AI systems would benefit from symbolic-connectionist approaches to improve user control. 

\subsubsection{Adaptive System Design for Different Design Stages}

In contrast to convergent processes, where participants desire systems to cater to their preferences (C2, C11), providing inspiration is needed in the divergent thinking stage. This is especially important, as images would easily lead to design fixation~\cite{cardoso2011influence, jansson1991design} (a drawback highlighted in Section~\ref{section:5.2}), and ``support from an AI image generator during ideation leads to higher fixation on an initial example''~\cite{wadinambiarachchi2024effects}. Previous research mentioned that clients may sometimes be uncertain about their own goals~\cite{horvitz1999principles}. Our study also revealed that client participants knew their true preferences better as they discovered \textbf{with} image AI. In the discovery process, encouraging divergent thinking is important, and surprises are favored~\cite{ko2023large}. In the divergent thinking stage, the system should encourage exploration and help broaden creative possibilities~\cite{weisz2023toward}, as exemplified by \textit{Luminate} that specifically caters to divergent thinking~\cite{suh2024luminate}. To address the distinct needs across stages, system designs may consider context-aware user interfaces, distinguishing diverging and converging processes, and providing context- and stage-aware system design.

\subsubsection{Providing Extra Assistance for Prompt Engineering and Collaborative Prompting}

Our findings suggest that prompt sharing benefits both stakeholders, and participants favored an integrated platform for collaborative prompting (C8, C11). This urged us to consider a collaborative prompting as a key feature to streamline their collective exploration. Apart from that, unlike deterministic tools such as computer-aided design and drafting (CADD) or sketching, which offer exceptional control, generative image AI operates more like a black box, and prompt engineering played a critical role. Refining prompts towards a design intent is arduous and takes many revisions~\cite{mahdavi2024ai}. Prompt templates (prompts with slots for people to fill in)~\cite{chang_prompt_2023}, and prompt lists (like Color, Material, Finish, etc.) would help broaden their prompting vocabulary. Existing research further suggests the need for an auto-revision prompt engineering tool--that can automatically correct and revise the structure of text prompts~\cite{ko2023large}. This feature is available in the \textit{Promptify} system, which can iteratively suggest changes to the user's original prompt~\cite{brade2023promptify}. Alternatively, conversational user interfaces (chatting with an agent and obtaining the desired design) offer more natural and intuitive interaction~\cite{vinker2024sketchagent}. These shed light on future directions for improving the prompting of generative image AI.

\subsubsection{Mitigating the Stylistic Bias in Image Generation}

Our participants observed that generation AI has strong stylistic preferences in its resulting images, as evidenced in the existing literature review~\cite{zhang2024research}. The innate biased nature of the humongous training dataset introduces a wide range of biases in the generative AI's outcomes~\cite{bender2021dangers, currie2024gender,drahl_2023_AI}. Our participants' remarks suggest that this was certainly at play. The data in the training set and the model we chose can bias the outcomes toward particular architectural styles and forms, negatively hindering the intent of communicating nuanced architectural styles, and resulting in reducing aesthetic diversity~\cite{manovich2018ai}. To mitigate the influence of bias, future systems could implement fairness measures in the AI model development stage, or explicitly reveal the biases within the dataset~\cite{de2023fair}. Alternatively, architectural companies or architects can fine-tune generative image models for their own architectural style~\cite{hu2021lora}. These approaches could enable generative image AI to better align with the user's communicative intent and capture their nuanced design style, fostering greater creative expression.




%% file: sections/7-conclusion.tex
\section{Conclusion}
Design communication has been a long-standing challenge in early-stage design processes, while generative image AI provides a promising communication medium. We experimented with generative image AI-mediated design communication scenarios involving early-career architect participants and client participants with a defined theme. The findings suggest that generative image AI can support more efficient design communication. By providing visual reference images that combine desired form and style in a high-fidelity format, generative image AI enables the externalization and representation of clients' preferences, with the speed of image generation affording rapid idea exchange and negotiation. This positively shifts the dynamics of design concept creation, benefiting client-architect relationships by more effectively eliciting the client's needs and helping to form a shared understanding. Limitations posed by generative AI, including the image generation biased results, can be further addressed. Our study suggests four directions for future system design for ethically utilizing generative image AI.

%% file: sections/z-appendix.tex
\section{Appendix}

\subsection{Recruitment Questionnaire}
\label{appendix:recruitmentquestionnaire}

The study aims to understand how designers use generative text-to-image AI tools to assist design communication. We invite designers from various design domains and anyone who wishes to serve as a client for a design project to join our study. Each study lasts 80 minutes. We will compensate you with a \$15 gift card for your time.

\begin{itemize}
    \item \textbf{Q1:} What kind of design work do you do? Or, do you want to be the client that someone designs for you?
     \\ O Architecture design \\ O Graphic design \\ O Product design
 \\ O Interior design \\ O Client (someone designs for you)
 \\ O Other \underline{\hspace{1cm}}
    
\item \textbf{Q2:} Are you a native English speaker? Or, do you match the following English testing requirements?
\\ O I am a native English speaker or have bilingual fluency in English.
\\ O I scored above the TOEFL 102, IELTS 7.5, or Duolingo 120 criteria.
\\ O No, I don't match the above criteria.

\item \textbf{Q3:} How many years have you studied and worked in your design domain in an English-speaking country? 
\\O 1-3 years \\O 4-5 years \\O 6+ years \\O I wish to be the client (that someone designs for you) \\ O Other (please specify)

\item \textbf{Q4: } Please provide your email if you are interested in joining our study. \\
\underline{\hspace{2cm}}

\end{itemize}

\subsection{On-boarding document}
\label{appendix:onboardingdocument}

\subsubsection{Text-to-image generation}

\begin{enumerate}
    \item Click  link (\url{https://beta.dreamstudio.ai/dream}) to open DreamStudio.
    \item Enter the prompt into the text box at the bottom of the page and click the Dream button to generate images. \footnote{When we conducted the study, the user interface changed, but the main features (like changing parameters) remained the same}
\end{enumerate}

\begin{figure}[h]
  \centering
  \includegraphics[width=0.5\linewidth]{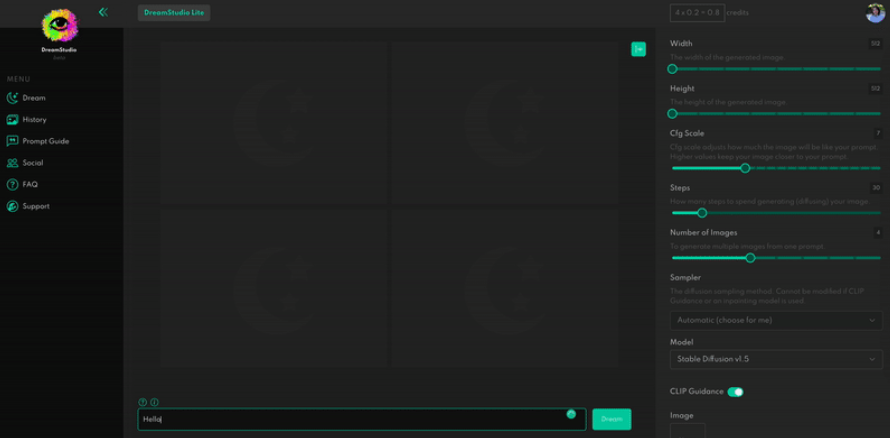}
  \Description{The left Figma board consists of black squares and text ``prompt'' beneath each black square. The right Figma board consists of many images filled inside the square, and beneath each image, there is a long text description.}
  \caption{DreamStudio user interface for image generation}
  \label{fig:example}
\end{figure}

\begin{table}[htp!]
\centering
\caption{Example prompts and corresponding images with different levels of detail}
\begin{tabular}{|m{3cm}|m{6.5cm}|m{2.4cm}|}
\hline
\textbf{Levels of Detail} & \textbf{Example Prompt} & \textbf{Generated Image} \\ \hline
Few words to describe & A modern kitchen & \includegraphics[width=2.3cm]{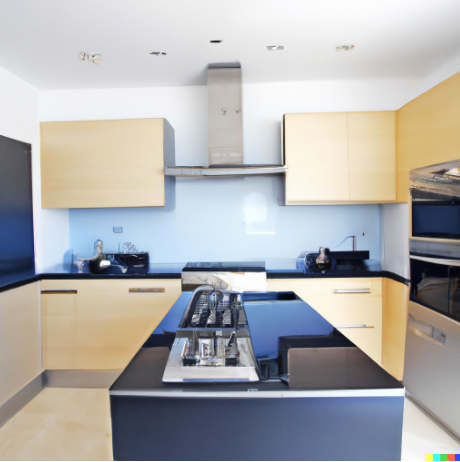} \\ \hline
Detailed descriptions & An Instagram-style modern design kitchen & \includegraphics[width=2.3cm]{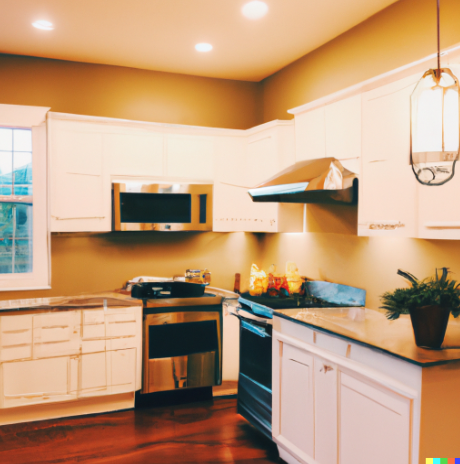} \\ \hline
More detailed description & 
\begin{tabular}[c]{@{}l@{}} 
An Instagram-style modern design kitchen with \\ 
gold fixtures, dark blue cabinets, white tiles, \\ 
and a farmhouse sink; professional picture. 
\end{tabular} & \includegraphics[width=2.3cm]{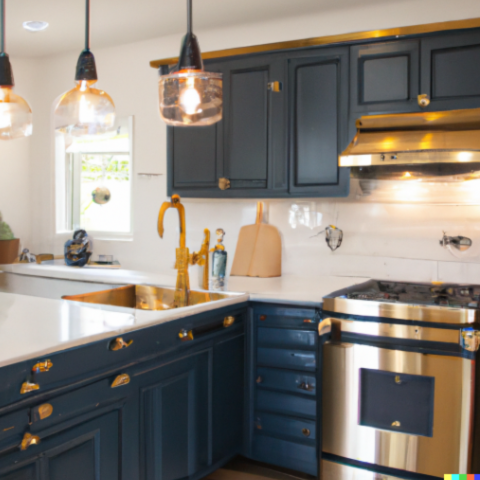} \\ \hline
\end{tabular}
\label{tab:example_prompts}
\end{table}

\section{Interview Questions}
\label{appendix:interviewquestions}

\begin{table}[H]
\centering
\caption{Interview Questions for Exploring AI Tool Integration in Design Processes}
\label{tab:interview_questions}
\renewcommand{\arraystretch}{1.2}
\begin{tabularx}{\textwidth}{@{}l>{\raggedright\arraybackslash}p{2.4cm}>{\raggedright\arraybackslash}X@{}}
\toprule
\textbf{No.} & \textbf{Topic} & \textbf{Interview Question} \\ \midrule

Q1 & The Experience &
1.1 Do you think the AI tool worked well? \newline
1.2 (If yes) Which parts of the AI tool worked well? \\ \midrule

Q2 & The Challenges &
2.1 Did you encounter any challenges using this tool? \newline
2.2 (If yes) Which parts did not work well? Tell me about the challenges you encountered. \\ \midrule

Q3 & The Improvement &
Based on the challenges you encountered, how could AI tools be improved to better support your design process? \\ \midrule

Q4 & Using the Tool in the Workflow &
4.1 (To client participants) Imagine you are communicating your design needs to a designer. Would you like to use the AI tool to facilitate design communication? \newline
(To architect participants) Imagine you are communicating your design to a client. Would you like to use the AI tool to facilitate design communication? \newline
4.2 (If yes) Do you think it would improve design communication or make it worse? (If no) Why not? Can you explain your reasoning? \\ \midrule

Q5 & Sharing the Prompt &
5.1 Did you use each other's prompts? \newline
5.2 Did sharing prompts help your design process? Why or why not? \\ \bottomrule
\end{tabularx}
\end{table}